\documentclass{article}
\usepackage[final]{neurips_2019}
\usepackage[percent]{overpic}
\usepackage[utf8]{inputenc}
\usepackage[T1]{fontenc}
\usepackage{threeparttable}
\usepackage{microtype}
\usepackage{hyperref}
\usepackage{booktabs}
\usepackage{amsfonts}
\usepackage{nicefrac}
\usepackage{graphicx}
\usepackage{titlesec}
\usepackage{authblk}
\usepackage{wrapfig}
\usepackage{xcolor}
\usepackage{url}

\definecolor{mymagenta}{rgb}{0.75,0,0.75}
\titlespacing{\section}{0pc}{0.02pc}{0pc}
\titlespacing{\subsection}{0pc}{0.02pc}{0pc}
\setcitestyle{numbers,square}

\title{Exploring Multi-Banking Customer-to-Customer Relations in AML Context with Poincaré Embeddings}

\author[1]{Lucia Larise Stavarache}
\author[2]{Donatas Narbutis}
\author[3]{Toyotaro Suzumura}
\author[1]{\authorcr Ray Harishankar}
\author[2]{Augustas {\v Z}altauskas}
\affil[1]{IBM Global Business Services}
\affil[2]{IBM Lithuania, Client Innovation Center Baltic}
\affil[3]{IBM T.J. Watson Research Center}

\begin{document}

\maketitle

\begin{abstract}
In the recent years money laundering schemes have grown in complexity and speed of realization, affecting financial institutions and millions of customers globally. Strengthened privacy policies, along with in-country regulations, make it hard for banks to inner- and cross-share, and report suspicious activities for the AML (Anti-Money Laundering) measures. Existing topologies and models for AML analysis and information sharing are subject to major limitations, such as compliance with regulatory constraints, extended infrastructure to run high-computation algorithms, data quality and span, proving cumbersome and costly to execute, federate, and interpret. This paper proposes a new topology for exploring multi-banking customer social relations in AML context -- customer-to-customer, customer-to-transaction, and transaction-to-transaction -- using a 3D modeling topological algebra formulated through Poincaré embeddings.
\end{abstract}

\section{Introduction}
AML (Anti-Money Laundering) exploration can take different pathways for each financial or related institution and there are multiple factors which contribute, e.g., institution type, segment, region, language, country regulations. Absence of a unified strategy alongside with known technical limitations present main vulnerability and expose such fragmented financial domain to continuously evolving fraud schema risks. KYC (Know Your Customer) and AML requirements represent old problems for the financial domain, however, in the recent years the increased strength of the rules for data privacy and security boosted them as the focus of each financial player with major impact and massive financial fines.

The impact of GDPR (General Data Protection Regulation) \cite{gdpr} in the financial field made it almost impossible for financial institutions to share malicious customer activity reports due to personal information constraints. The emphasized data privacy measures that had to be implemented to protect genuine customers also allow for the wicked ones to thrive -- situation in many aspects resembles the ``chicken and egg'' paradigm \cite{chicken-egg}. Groups of malicious customers usually operate as networks across multiple banks, therefore, analysis of their connectivity has to be performed.

Current technology solutions explore various topological structures, e.g., matrices \cite{aml-solutions}, graphs \cite{DBLP:journals/corr/Le-KhacMOBK16}, and various variants of neural networks \cite{DBLP:journals/corr/abs-1812-00076} with known limitations: the amount of data that can be analyzed, its quality and duplication, possibility to discover logical and cross-referenced relationships that can point to SAR (Suspicious Activity Reports). Given the current topologies, the complexity of each structure can reach $O(n^2)$ limiting the scaling, while transaction-to-transaction scenarios or correlating customer-to-transactions in billion-row data volumes becomes impossible to explore.

Recent year breakthroughs shifted the perspective from relational matrix analysis on a big data systems (e.g. Hadoop) to applications of homogeneous graphs and neural networks, which enhance discovery of customer relationships and their interaction patterns. However, scalability of these models is limited by a large memory use, slow learning rate, difficulty to transfer them to new datasets, and low capability for distributed architectures. In banking ecosystems restricted access to information and business flows makes the real-time processing difficult. Therefore, capturing SAR patterns and transferring them to an exhaustive topology that can learn, share, and express features in a federated multi-banking manner, as proposed by this paper, addresses a major industry problem.

The KYC/AML ecosystem involves multiple parties and steps for the detection of SARs, implying that analysts and regulators must have the ability to validate, explore, make corrections, feed new findings/patterns into the system, and have historical explainability of the results. Reporting back to the mentioned constraints, this paper proposes a new approach and topology for expressing the customer relationships, allowing to visualize and explore the data in a spherical 3D Poincaré space. The approach accounts for transformation from a sparse matrix 2D into a finite condensed 3D space, i.e., matrices are $O(n^2)$ and high-graphs $O(n\log(n))$. Replacing the Euclidean distances with Poincaré allows the 3D sphere to be limited to a finite radius and create a natural taxonomy classifier of the customer relationships.

\section{Data and Approach}
The dataset was designed by the Financial Conduct Authority (UK) for their annual ``2019 Global AML and Financial Crime TechSprint'' \cite{techsprint2019} in collaboration with SYNTHETICR, considering real world scenarios and AML/KYC patterns, i.e., laundromat, layered, dispersal, and collecting networks, among others. The explored dataset is based on a simulated multi-banking transaction scenario, comprising of six individual financial institutions with the characteristics provided in Table~\ref{table-corpora}.

\begin{wraptable}{l}{7.6cm}
\vspace{-7mm}
\caption{Data corpora characteristics of six banks.}
\label{table-corpora}
\vspace{8mm}
\centering
\begin{scriptsize}
\begin{tabular}{ll}
\toprule
Corpora component & Volumetric \\
\midrule
Accounts per bank & 1. 273k -- large \\
 & 2. 177k -- medium \\
 & 3. 154k -- medium \\
 & 4. 147k -- medium \\
 & 5. 95k -- small \\
 & 6. 74k -- small \\
\midrule
Account holders & -- 396k -- female \\
 & -- 316k -- male \\
 & -- 58k -- company \\
\midrule
Unique individuals & $\sim$200k after entity resolution \\
 & typo corrections and deduplication \\
\midrule
Customer nationality & -- 750k UK \\
 & -- 11k Poland \\
 & -- 5k Romania \\
 & -- 4k India \\
 & -- others \\
\bottomrule
\end{tabular}
\end{scriptsize}
\vspace{-2mm}
\end{wraptable}

The total amount of data provided in CSV format was $\sim$250 GB, containing $\sim$0.9 million accounts and $\sim$1 billion transactions over period of 24 months distributed across six banks. The CSV files were converted into HDF5 format using Vaex \cite{vaex} for real-time scanning, e.g., selecting all transactions of a target account and displaying transaction money amount as a function of time. The analyzed corpora contained only $\sim$10\% (108k) transactions in the dataset, which were provided with ``source-origin'' and ``destination-origin'' account numbers, i.e., not cash, contact-less, or open-ended operations. For fast interactive exploratory analysis and visualization TOPCAT tool \cite{topcat} was used.

The dataset contained mappings of the following relationships: customer-to-bank, bank-to-account, account-to-transaction, customer-to-customer, customer-to-related-party, related-party-to-risk-intelligence. Data tables used for analysis and their aliases are presented in Table~\ref{table-schemas}; columns {\tt\small doc\_id}, {\tt\small entity\_id}, and {\tt\small relation\_duration\_months} were created. The financial crime activity for a specific customer was marked in {\tt\small RISK} table, column {\tt\small fincrime\_risk\_exit} was used as main indicator, while columns {\tt\small black\_list} and {\tt\small aml\_flag} were used as secondary indicators of suspicious activities.
\newline

\begin{table}
\caption{Data table schemas used for analysis.}
\label{table-schemas}
\centering
\begin{scriptsize}
\begin{tabular}{llll}
\toprule
Customer Profile & Customer Related Party Link & Customer Related Parties & Party Risk Intelligence \\
\midrule
Columns: 41 & 7 & 26 & 9 \\
\midrule
Alias: {\tt\scriptsize CUSTOMERS} & {\tt\scriptsize LINK} & {\tt\scriptsize PARTIES} & {\tt\scriptsize RISK} \\
\midrule
{\tt\scriptsize bank\_id} & {\tt\scriptsize bank\_id} & {\tt\scriptsize bank\_id} & {\tt\scriptsize bank\_id} \\
{\tt\scriptsize bank\_name} & {\tt\scriptsize bank\_name} & {\tt\scriptsize bank\_name} & {\tt\scriptsize bank\_name} \\
{\tt\scriptsize customer\_id} & {\tt\scriptsize customer\_id} & {\tt\scriptsize customer\_id} & {\tt\scriptsize customer\_id} \\
{\tt\scriptsize id\_doc\_number} & ... & {\tt\scriptsize id\_doc\_number} & ... \\
{\tt\scriptsize company\_registration\_id} & ... & {\tt\scriptsize company\_registration\_id} & ... \\
{\tt\scriptsize doc\_id} & ... & {\tt\scriptsize doc\_id} & ... \\
... & {\tt\scriptsize related\_party\_id} & {\tt\scriptsize related\_party\_id} & {\tt\scriptsize related\_party\_id} \\
... & {\tt\scriptsize relation\_start\_date} & ... & ... \\
... & {\tt\scriptsize relation\_end\_date} & ... & ... \\
... & ... & ... & {\tt\scriptsize fincrime\_risk\_exit} \\
{\tt\scriptsize entity\_id} & {\tt\scriptsize relation\_duration\_months} & {\tt\scriptsize entity\_id} & ... \\
\bottomrule
\end{tabular}
\vspace{0.5mm}
\begin{tablenotes}
\item {\footnotesize \quad {\tt\footnotesize CUSTOMERS} and {\tt\footnotesize RISK}: 923,622 rows. {\tt\footnotesize fincrime\_risk\_exit} for 2,785 banned customers.}
\item {\footnotesize \quad {\tt\footnotesize id\_doc\_number}: 90\% (individuals). {\tt\footnotesize company\_registration\_id}: 10\% (companies).}
\end{tablenotes}
\end{scriptsize}
\vspace{-6mm}
\end{table}

Two money laundering schemes ``collecting network'' and ``layered network'' were investigated on the dataset in preparation for the Poincaré embedding of customer-to-customer relationship mapping to visualize fraudulent relationships in 3D space. The ``collecting network'' describes customers that act as an entry point for distinct networks of criminal customers (individuals or companies). Accounts that were opened and closed within the time span of 8 months were considered, correlating them with customers associated to the accounts with {\tt\small fincrime\_risk\_exit = true}. The list of customers that most likely belong to this scheme was produced, identifying 124 account with accuracy of 90\%. 

The ``layered network'' activities are focused on the accounts that are connected to criminals. Any account that had any transfer with a confirmed criminal account identified in ``collecting network'' was considered. All the transactions per such account were binned into weeks with the underlying idea that this should be enough time to transfer money to and from the account. For each account every week the number of transactions and accumulated sum for inbound and outbound transfers was calculated. To take into consideration the low balance at the end of the week, all the accounts that had a balance larger than 20\% of the accumulated sum of inbound transactions were filtered. This resulted in 288 identified fraudulent accounts with an accuracy of 90\% on confirmed results.

However, the main analyzed relationship of this study was a customer-to-customer relation, comprising of social interactions, their duration time, activity, and risk criteria. It was considered to be the central relationship to explore under the following hypothesis: a customer (person identified by {\tt\small id\_doc\_number}, i.e., passport, or company identified by {\tt\small company\_registration\_id}, i.e., number given by common registry) can have numerous accounts in different banks and the linkage via customer-to-bank, bank-to-account, and account-to-transaction relations interconnects customers into a relation graph.

The graph can be imagined as arranged in three homogeneous planes (individuals, accounts, transactions) with the following properties: a) individuals can have 1:M accounts over multiple banks, b) some of individuals have associated risk intelligence records across different banks, c) account has 1:M transactions that identify behavior patterns by time, frequency, amount or operation. The above proposed 3D model of the customer relation graph extends enormously in Euclidean space. However, once embedded in $n$-dimensional Poincaré space (in subsequent analysis 3D is used), it is limited to a finite radius, allowing visualization, and has customer relationships classified by taxonomy. There is a possibility to extend the analysis in transaction time domain, since AML fraud schemes have patterns in time. From the provided data, a social relation graph of $\sim$200,000 unique individuals within a simulated multi-banking dataset using Poincaré embeddings was explored. Two types of customers were considered: individuals and companies. Links between customers based on their social relations and their duration time was considered.

{\it Note 1}. Using existing approaches this large graph could be represented as a sparse square matrix, which would require $200\times200$ typical computer screens arranged as a grid to display it. Visual inspection of relationship matrix was performed without obtaining insight on the sparsity pattern in representation of the data, except that the connection structure on the large scale view looks random.

{\it Note 2}. There is a significant interest in banking data analysis to tackle the anti-money laundering problem via transaction analysis. A graph learning approach has been introduced \cite{DBLP:journals/corr/abs-1812-00076} and was based on simulations within a single bank. Evolving graph convolutional networks have been proposed for node and edge classification, and link prediction \cite{DBLP:journals/corr/abs-1902-10191}.

{\it Note 3}. Big data manipulation exposes common ``knowns'' as incompleteness, redundancy, inconsistency and incorrectness, therefore entity resolution curation is a necessary step before starting the actual data analysis.

\section{Experiments}
Embeddings are a widely used approach to create a low-dimension representation of a given graph \cite{DBLP:journals/corr/abs-1806-07464}. Comprehensive surveys have been provided for graph \cite{DBLP:journals/corr/abs-1709-07604} and network \cite{DBLP:journals/corr/abs-1711-08752} embeddings. Poincaré embeddings have shown a good performance in learning the representations of graphs in $n$-dimensional Poincaré ball by capturing context and hierarchy of related entities \cite{DBLP:journals/corr/abs-1905-09791}. Taking as reference point the approach developed developed by \cite{DBLP:journals/corr/NickelK17} (see for the detailed description of the method and its application on various graphs), this paper explores the customer-to-customer relationship in a 3D Poincaré space, extending the approach into a banking domain with the purpose to identify potential suspicious/criminal activities across the banking operations.

\subsection{Customer links}
The linkages for the customer-to-customer relationship analysis were mapped using the data schema and table aliases described in Table~\ref{table-schemas}. Columns of the tables are referenced by table's name as a prefix, e.g., {\tt\small CUSTOMERS:bank\_id}. To define the Poincaré sphere space, the customer-to-customer relationship from table {\tt\small PARTIES} was modelled with the following data observations: {\tt\small PARTIES:entity\_id} contains only individuals, not companies. To identify unique customers an internal match (i.e. group by) on {\tt\small PARTIES:id\_doc\_number} was performed, resulting in 63,912 unique entities with sparse relationships links in range from 0 to 32 with various {\tt\small PARTIES:related\_party\_id} entities.

To measure relationship duration table {\tt\small LINK} was used to extract relationship start-end information. Relationship duration time ({\tt\small LINK:relation\_duration\_months}) is an attribute of the link and is used as a weight of the graph edge between the nodes (entities) when performing the Poincaré embedding. The longest relation duration was $\sim$600 months.

To identify entities in {\tt\small CUSTOMERS} and {\tt\small PARTIES} tables, {\tt\small id\_doc\_number} was used for individuals and {\tt\small company\_registration\_id} for companies. An entity is either an individual or a company, but no companies were found in {\tt\small PARTIES} as noted above. A new {\tt\small doc\_id} column was created in these tables from {\tt\small id\_doc\_number} with prefix ``individual\_'' or from {\tt\small company\_registration\_id} with prefix ``company\_''. Finally, internal match was performed on {\tt\small CUSTOMERS:doc\_id} and {\tt\small PARTIES:doc\_id} columns separately to count unique entities and to create {\tt\small CUSTOMERS:entity\_id} and {\tt\small PARTIES:entity\_id} columns. Entity cross-identification was realized through the following cross-joins:

\begin{flushleft}
\begin{small}
{\tt\small PL} = merge({\tt\small PARTIES}, {\tt\small LINK}, \\
\qquad left\_on=[{\tt\small PARTIES:bank\_id}, {\tt\small PARTIES:related\_party\_id}], \\
\qquad right\_on=[{\tt\small LINK:bank\_id}, {\tt\small LINK:related\_party\_id}])

{\tt\small PLC} = merge({\tt\small PL}, {\tt\small CUSTOMERS}, \\
\qquad left\_on=[{\tt\small LINK:bank\_id}, {\tt\small LINK:customer\_id}], \\
\qquad right\_on=[{\tt\small CUSTOMERS:bank\_id}, {\tt\small CUSTOMERS:customer\_id}])

{\tt\small PLCR} = merge({\tt\small PLC}, {\tt\small RISK}, \\
\qquad left\_on=[{\tt\small CUSTOMERS:bank\_id}, {\tt\small CUSTOMERS:customer\_id}], \\
\qquad right\_on=[{\tt\small RISK:bank\_id}, {\tt\small RISK:customer\_id}], how='left'))
\end{small}
\end{flushleft}

Further analysis of the resulting table {\tt\small PLCR} showed that the customer relation graph contains links from individuals to companies. To create a data structure suitable for Poincaré embeddings with \cite{DBLP:journals/corr/NickelK17} program, a list of tuples, consisting of ({\tt\small ID1}, {\tt\small ID2}, {\tt\small weight}) must be provided. Since {\tt\small PARTIES:entity\_id} can be linked to {\tt\small CUSTOMERS:entity\_id} via several banks (up to six) with different relation duration times, it was decided to treat these relations separately. Finally, the data created for Poincaré embeddings contained these columns: {\tt\small ID1} = {\tt\small PARTIES:entity\_id}, {\tt\small ID2} = {\tt\small CUSTOMERS:bank\_id} + {\tt\small CUSTOMERS:entity\_id}, {\tt weight} = {\tt\small LINK:relation\_duration\_months}. There are 205,207 edges (relation links) in the graph and 151,044 nodes -- unique entities, i.e., individuals and companies, which are treated as separate entities if have accounts in different banks.

\subsection{Poincaré relationship mapping}
A entity-to-entity relation graph was constructed from the list of (entity, entity, weight) tuples, which is passed as input to the algorithm to perform Poincaré embedding. The embedded entity (individual or a company) is represented as a 3D vector of float numbers within ranges $-1$ to 1. The entity representation vectors are moved in the Poincaré space using the following transformation sequence: a) initially the vectors are assigned random values close to zero (Gaussian with zero mean and standard deviation of 0.001), b) in each step a random pairing of entities is performed, c) for linked entities their vectors are brought closer together and not linked entities the vectors are moved further apart by computing the objective function as a sum of Poincaré distances between the vectors as described in \cite{DBLP:journals/corr/NickelK17}.

\begin{figure}
\centering
\begin{overpic}[trim=0.8cm 0.8cm 0.8cm 1.4cm,clip,width=0.48\textwidth,natwidth=800,natheight=800]{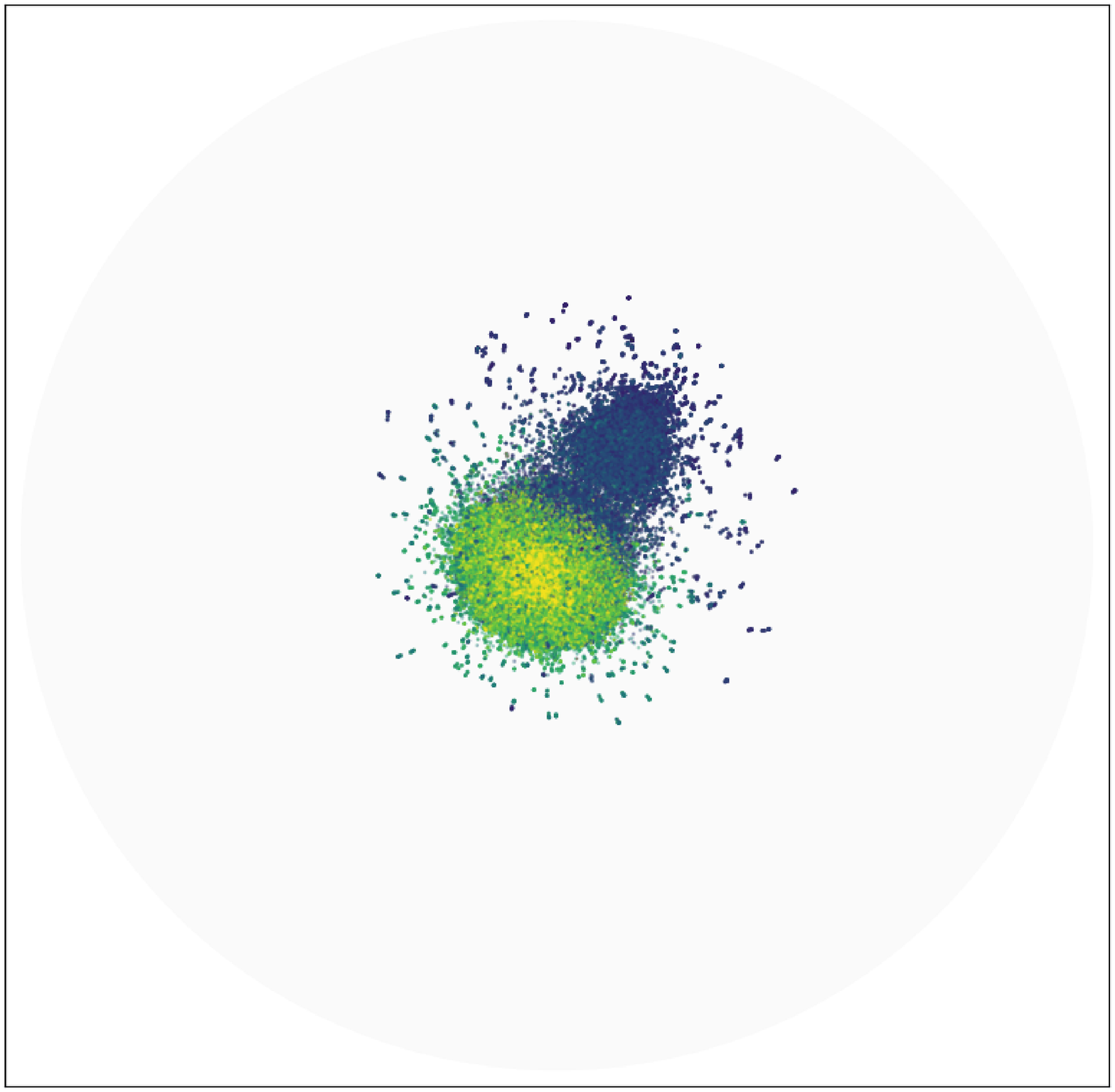} \put(0,84){a)} \put(0,94){Iteration: 30}
\put(49.5,47){\color{black}\vector(1,0){35}}
\put(49.5,47){\color{black}\vector(0,1){35}}
\put(49.5,47){\color{black}\vector(-1,-1){19}}
\end{overpic}
\begin{overpic}[trim=0.8cm 0.8cm 0.8cm 1.4cm,clip,width=0.48\textwidth,natwidth=800,natheight=800]{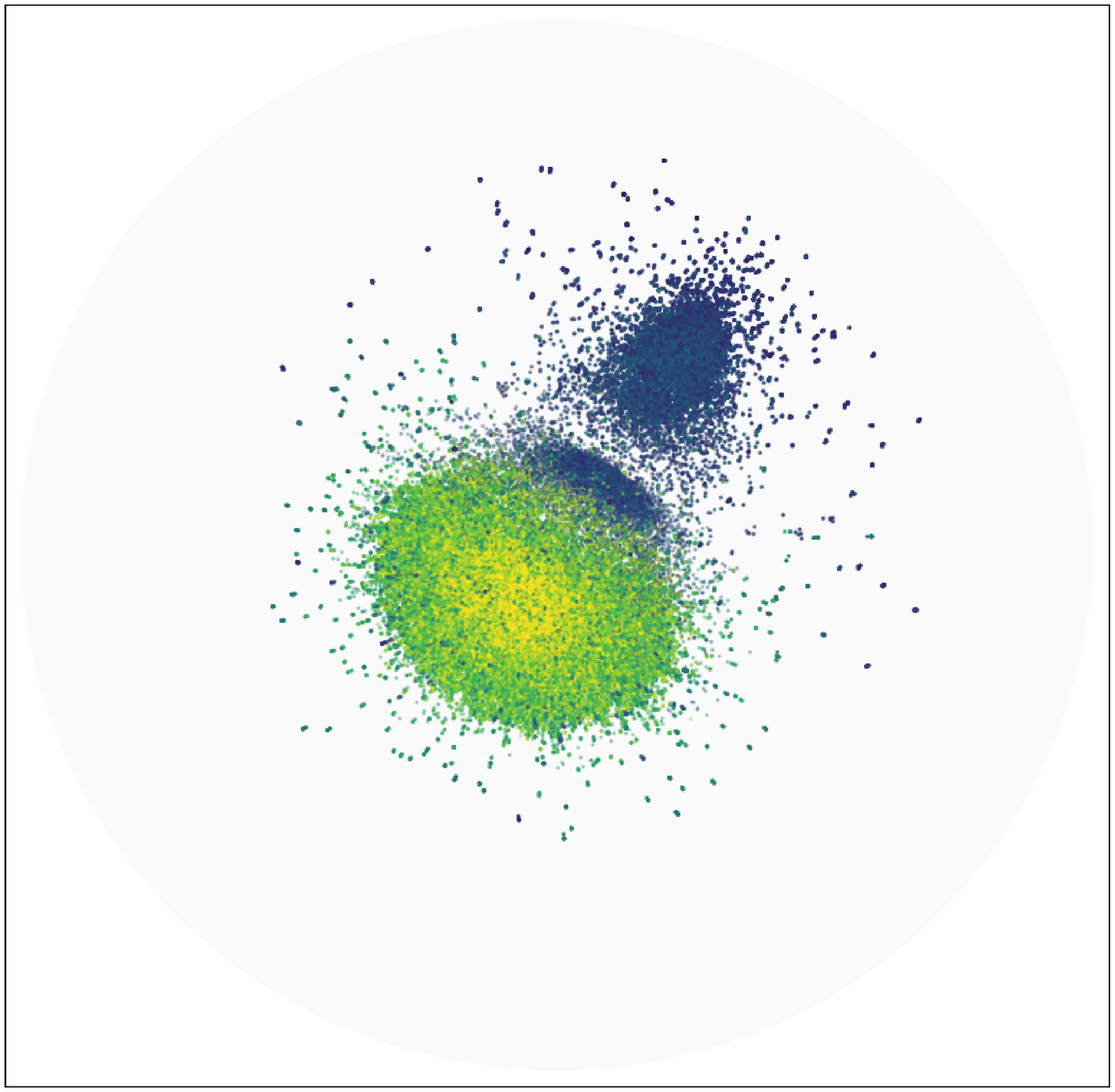} \put(0,84){b)} \put(0,94){Iteration: 40} \end{overpic}
\begin{overpic}[trim=0.8cm 0.8cm 0.8cm 1.4cm,clip,width=0.48\textwidth,natwidth=800,natheight=800]{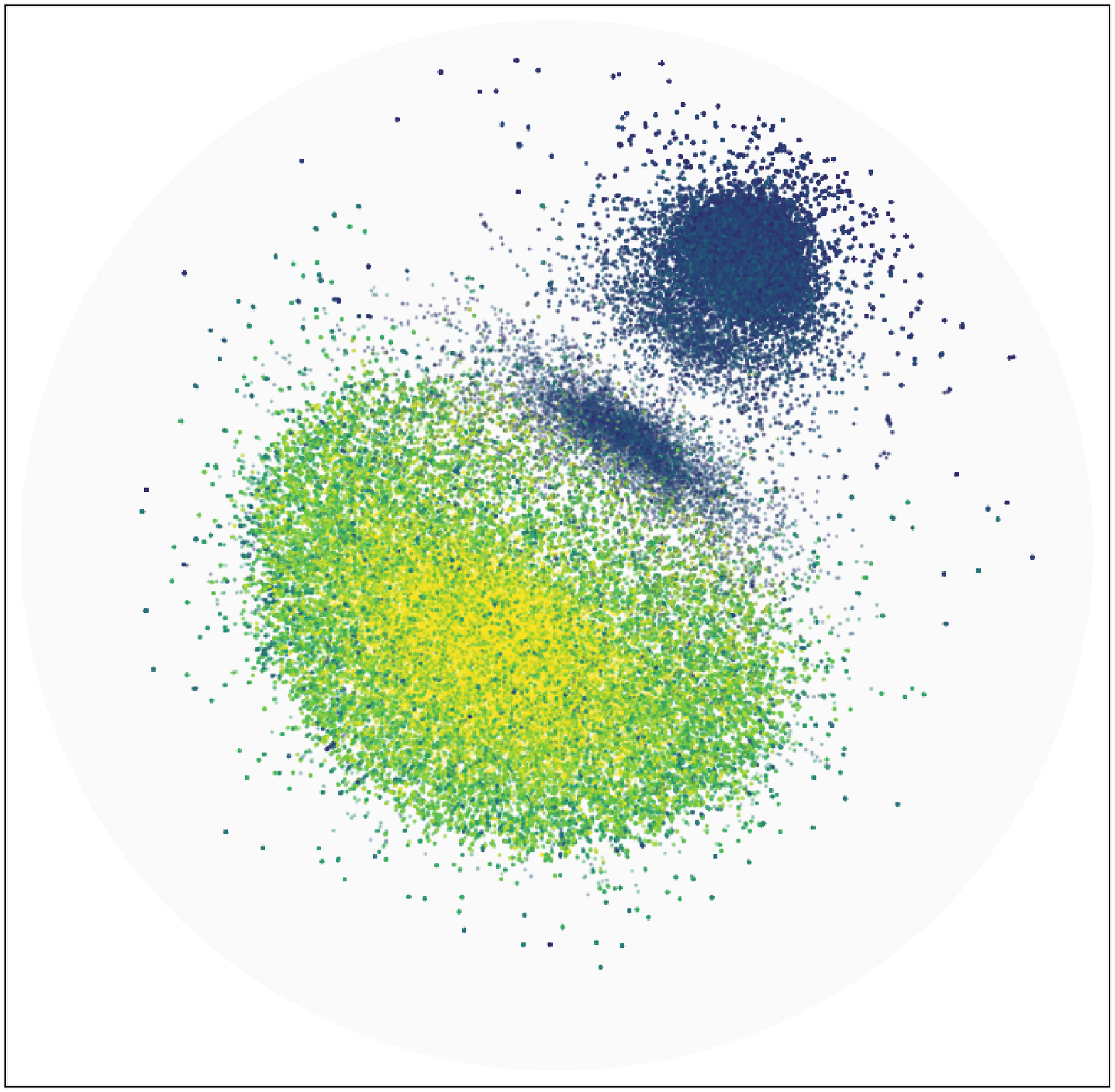} \put(0,84){c)} \put(0,94){Iteration: 60} \end{overpic}
\begin{overpic}[trim=0.8cm 0.8cm 0.8cm 1.4cm,clip,width=0.48\textwidth,natwidth=800,natheight=800]{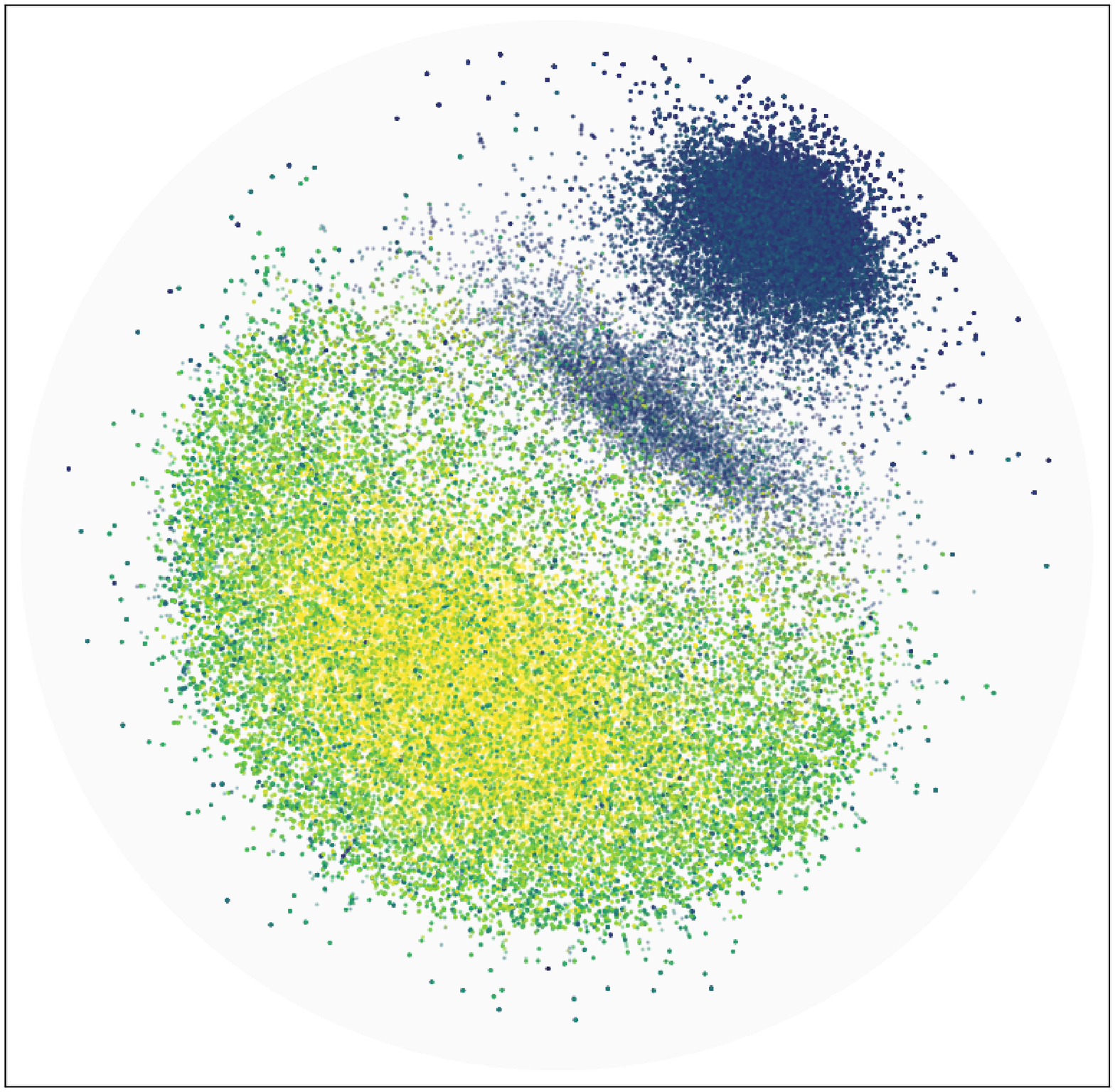} \put(0,84){d)} \put(0,94){Iteration: 80} \end{overpic}
\caption{3D Poincaré embeddings of the customer-to-customer relations are projected onto a plane. Dots show entity vectors with origin at the center, color-coded corresponding to number of connections they have, from 2 (yellow/green) to more than 10 (dark blue). Grey shaded circle is a Poincaré sphere of radius 1. Panels show progress of embedding iterations 30, 40, 60, and 80. Three groupings of customers emerge in panel b).}
\label{figure-learn}
\end{figure}

Fig.~\ref{figure-learn} shows sequence of embeddings representations during the progress of model training; number of training iterations was limited due to data access policy set by the ``2019 Global AML and Financial Crime TechSprint'' \cite{techsprint2019} and additional iterations could have improved the results.

As can be seen in Fig.~\ref{figure-learn}, it takes 80 iterations for some entity embedding vectors to asymptotically approach length of $\sim$1 in Euclidean space, which corresponds to infinite distance in Poincaré space.

A useful feature of Poincaré representation is that the maximum length of embedding vector is 1, and since it is a non-linear space, entity vector representations are approaching 1 asymptotically. If the entities were not connected, they would be embedded on the surface of radius = 1 sphere and with maximum separation along great circle lines connecting them. If entities are highly connected, they are brought closer together.

As can be seen in Fig.~\ref{figure-learn}, Poincaré embeddings of customer-to-customer relationship revealed interesting groupings of entities. Entities with low and high connectivity are initially separated into two clusters (Fig.~\ref{figure-learn} b). Low connectivity entities are pushed towards the boundary of the lower left half of the sphere. Upper right is occupied by $\sim$28,000 high connectivity entities. As training progresses, an intermediate elongated group appears (Fig.~\ref{figure-learn} c--d), containing $\sim$6,000 entities.

Although the initial embedding vector values and the training process is random, by running the embeddings with different initial conditions it was verified that the clusters finally obtained are stable and can be reliably distinguished from a random chance. Only the 3D rotation of the whole system is arbitrary.

\section{Results and Discussion}
\begin{figure}
\centering
\begin{overpic}[trim=0.8cm 0.8cm 0.8cm 1.4cm,clip,width=0.48\textwidth,natwidth=800,natheight=800]{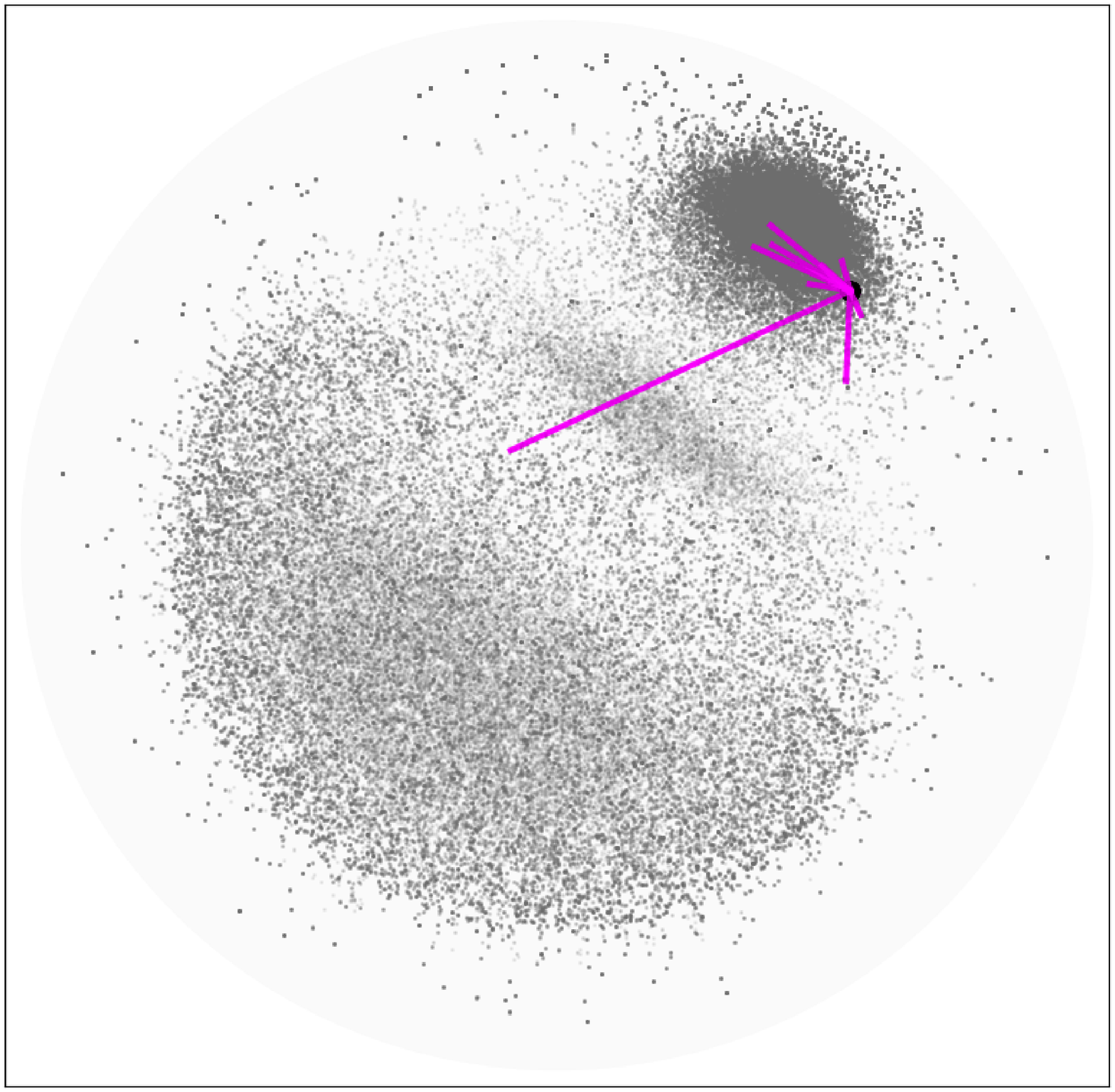} \put(0,84){a)} \put(0,94){\color{mymagenta} Suspicious: 10} \end{overpic}
\begin{overpic}[trim=0.8cm 0.8cm 0.8cm 1.4cm,clip,width=0.48\textwidth,natwidth=800,natheight=800]{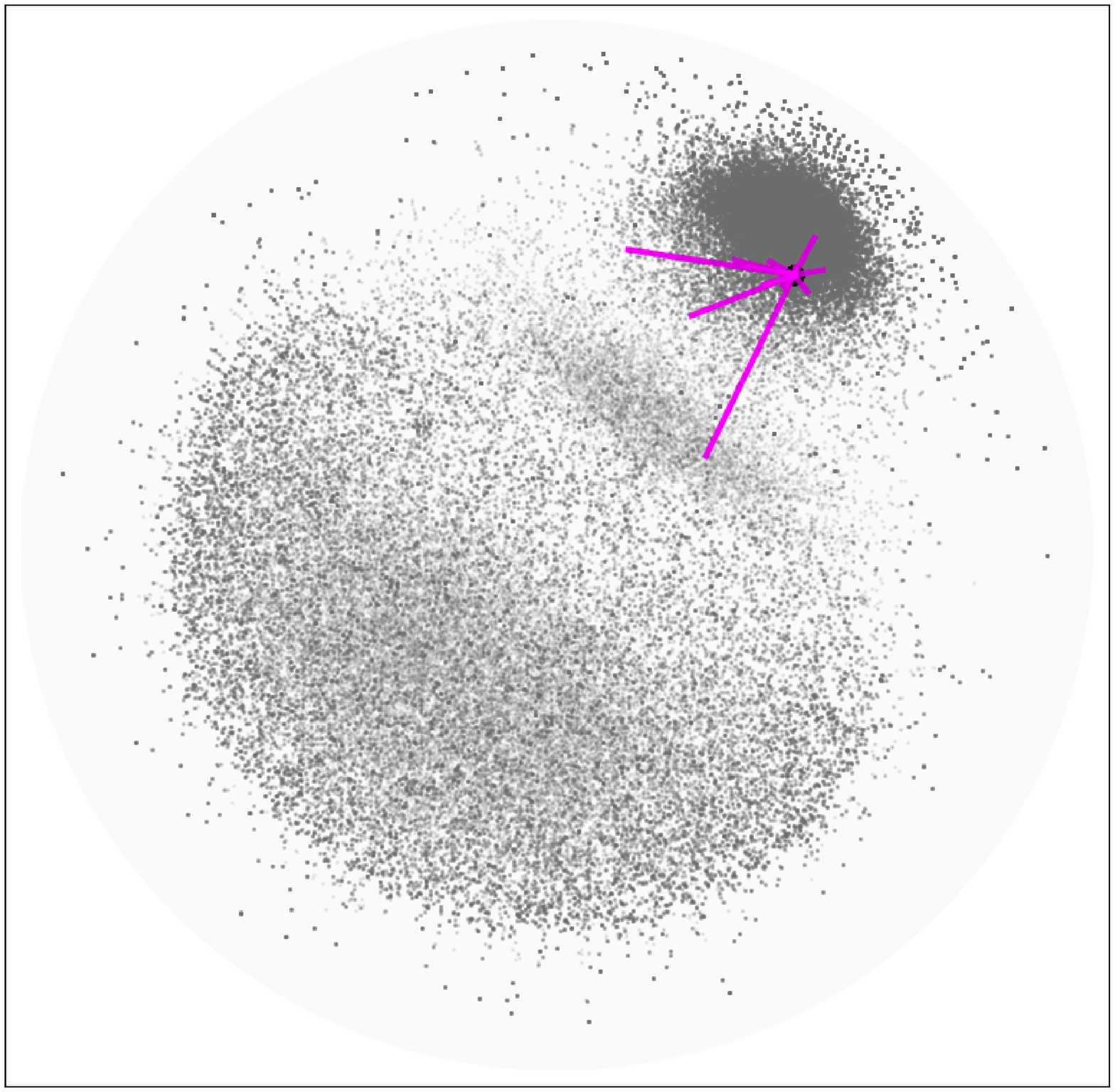} \put(0,84){b)} \put(0,94){\color{mymagenta} Suspicious: 11} \end{overpic}
\begin{overpic}[trim=0.8cm 0.8cm 0.8cm 1.4cm,clip,width=0.48\textwidth,natwidth=800,natheight=800]{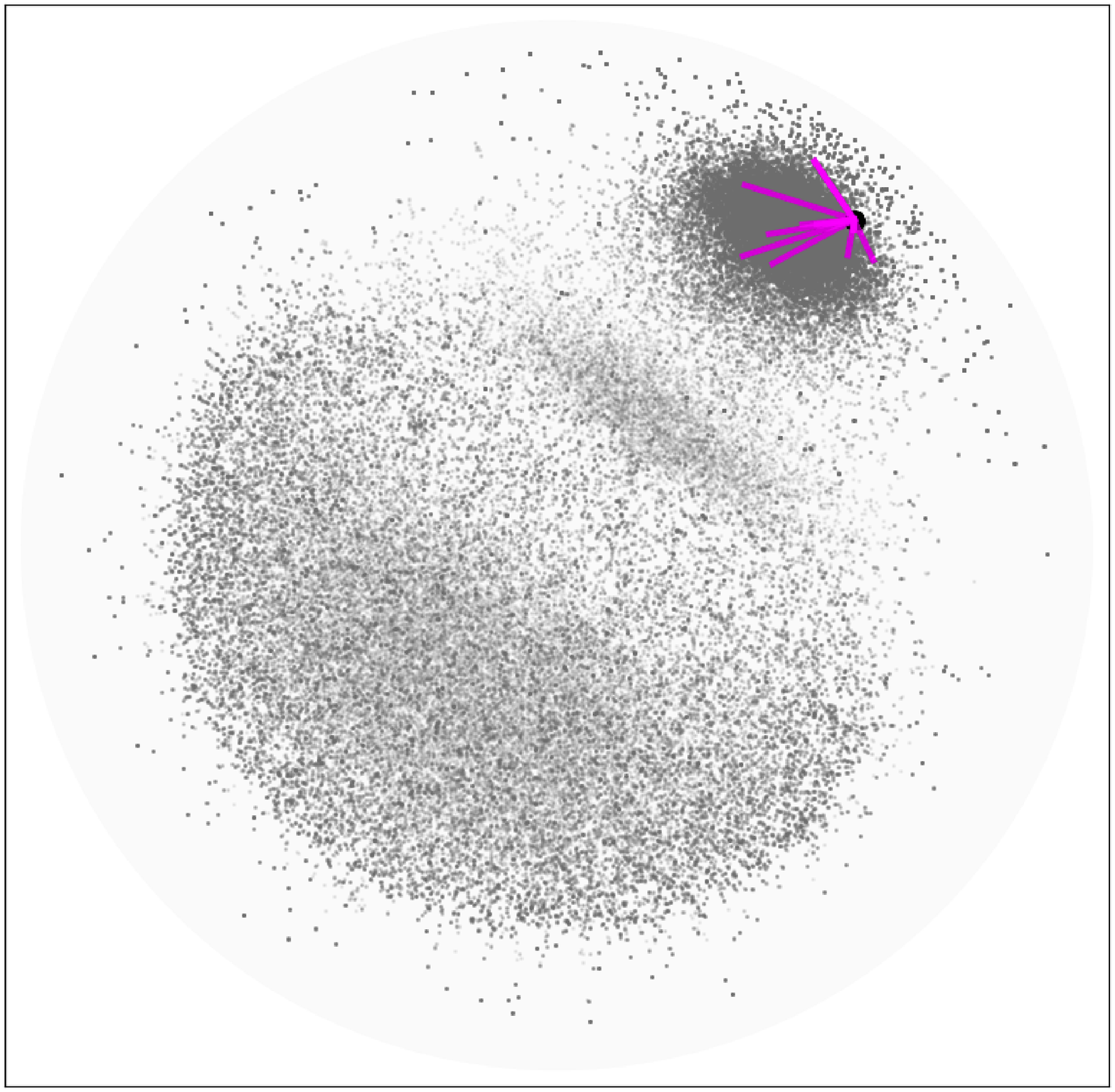} \put(0,84){c)} \put(0,94){\color{mymagenta} Suspicious: 14} \end{overpic}
\begin{overpic}[trim=0.8cm 0.8cm 0.8cm 1.4cm,clip,width=0.48\textwidth,natwidth=800,natheight=800]{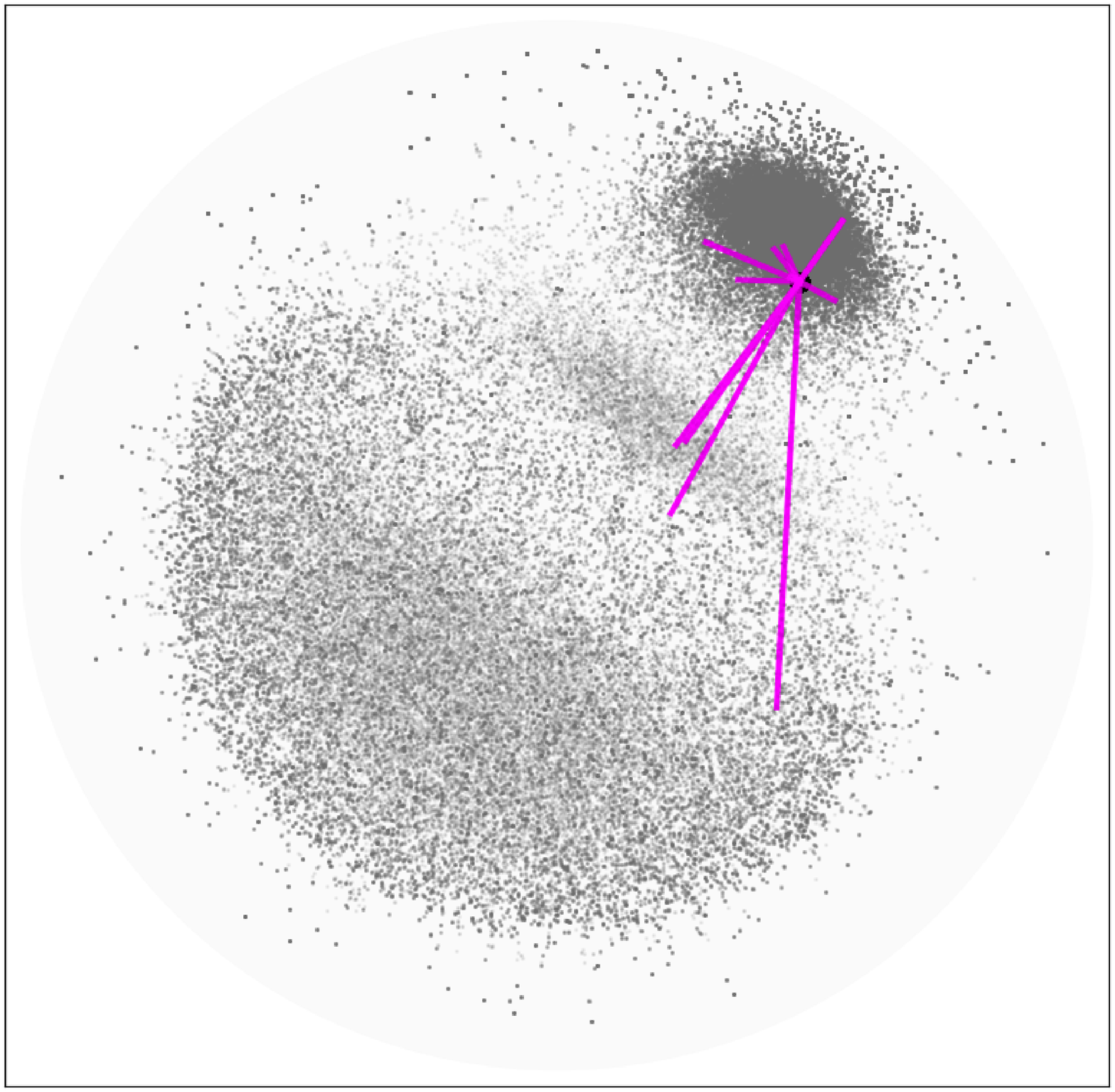} \put(0,84){d)} \put(0,94){\color{mymagenta} Suspicious: 15} \end{overpic}
\caption{Same as Fig.~\ref{figure-learn} d) after 80 iterations of Poincaré embedding, but color-coded in gray. Each panel shows a single entity marked by positive {\tt\small fincrime\_risk\_exit} and magenta lines linking it to its related parties. Number of links is indicated in top left corner of each panel.}
\label{figure-crime}
\end{figure}

Fig.~\ref{figure-crime} shows that entities marked with positive {\tt\small fincrime\_risk\_exit}, i.e., have been banned from banks due to financial crimes, are highly-connected and belong to the grouping of other highly-connected entities. Each panel shows the ``suspicious'' entity with magenta lines indicating links to its related parties. Majority of the ``suspicious'' entities reside in the upper right compact grouping and have connections mostly within this grouping or intermediate elongated grouping below. Note, that connections are displayed as lines assuming Euclidean space for simplicity.

\begin{figure}
\centering
\begin{overpic}[trim=0.8cm 0.8cm 0.8cm 1.4cm,clip,width=0.48\textwidth,natwidth=800,natheight=800]{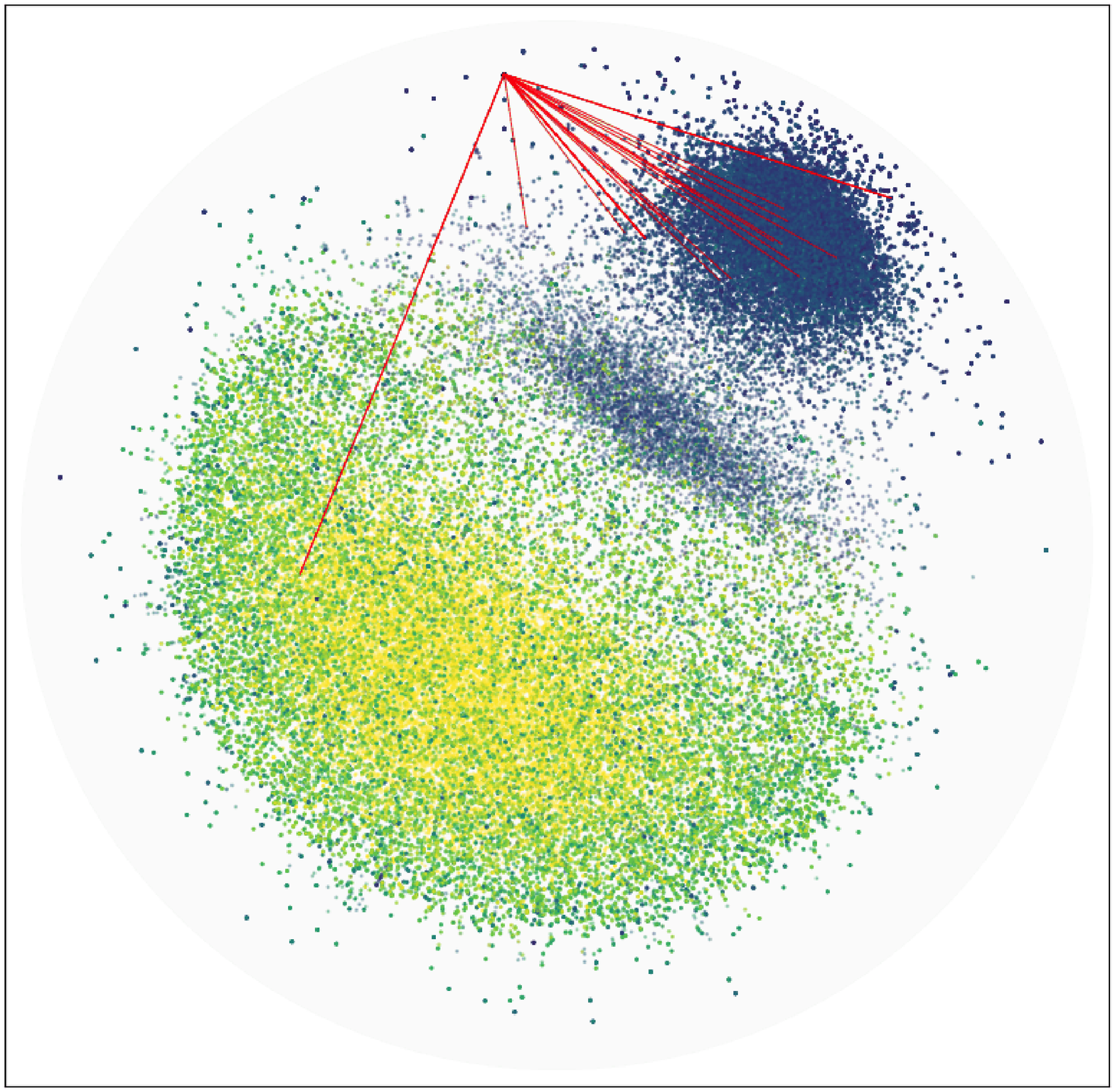} \put(0,84){a)} \put(0,94){\color{red} Links: 32} \end{overpic}
\begin{overpic}[trim=0.8cm 0.8cm 0.8cm 1.4cm,clip,width=0.48\textwidth,natwidth=800,natheight=800]{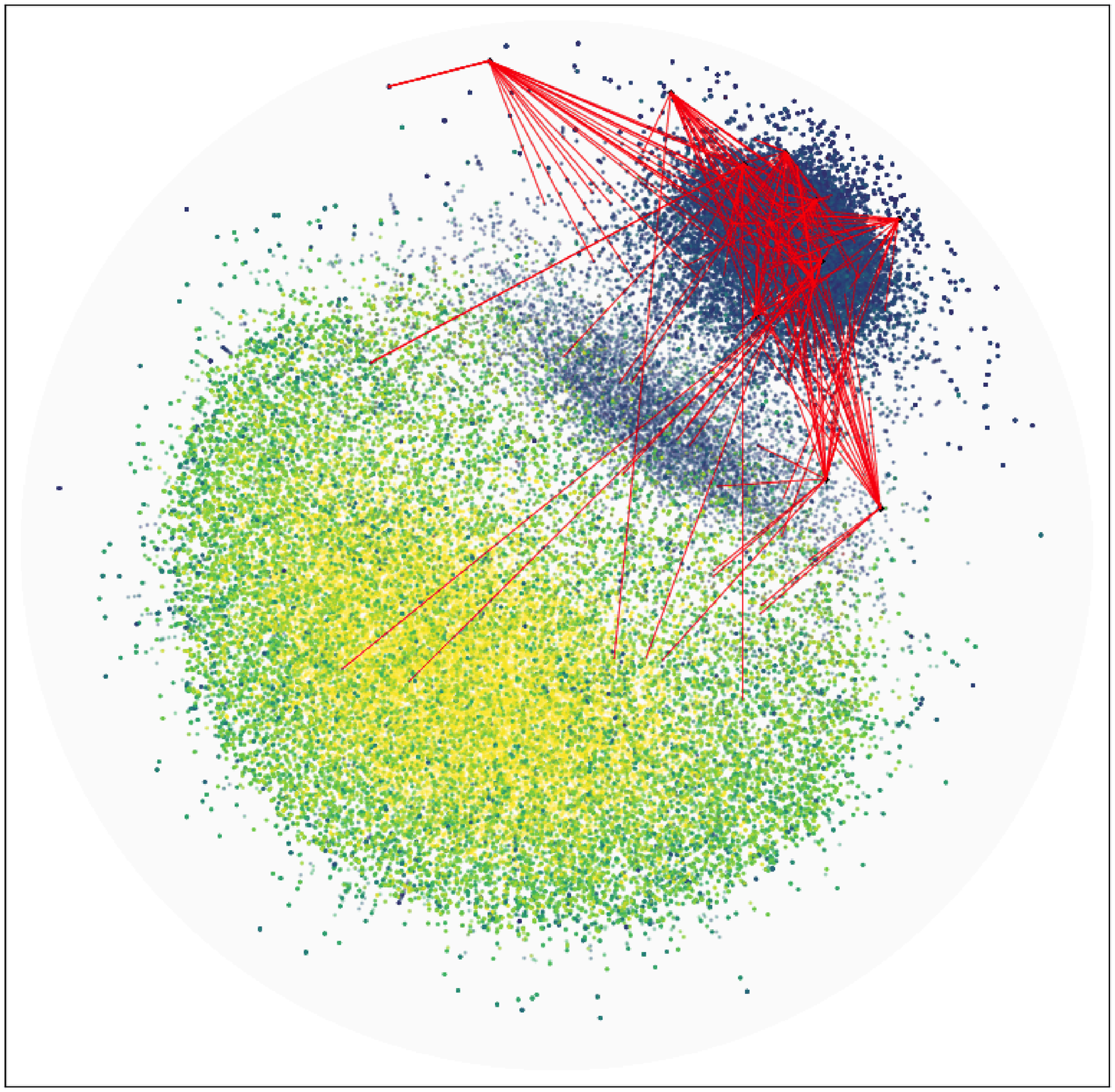} \put(0,84){b)} \put(0,94){\color{red} Links: 25} \end{overpic}
\begin{overpic}[trim=0.8cm 0.8cm 0.8cm 1.4cm,clip,width=0.48\textwidth,natwidth=800,natheight=800]{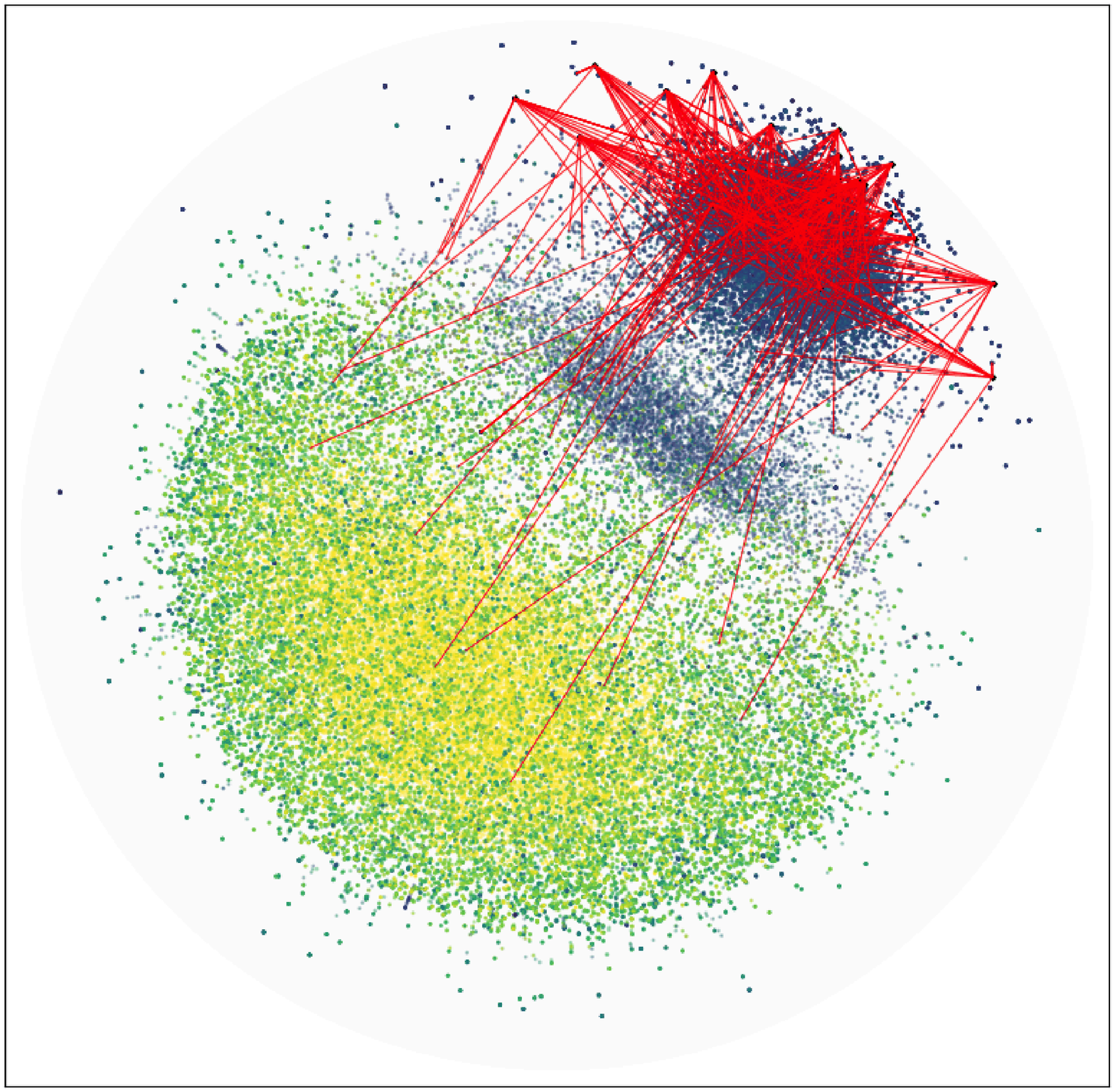} \put(0,84){c)} \put(0,94){\color{red} Links: 23} \end{overpic}
\begin{overpic}[trim=0.8cm 0.8cm 0.8cm 1.4cm,clip,width=0.48\textwidth,natwidth=800,natheight=800]{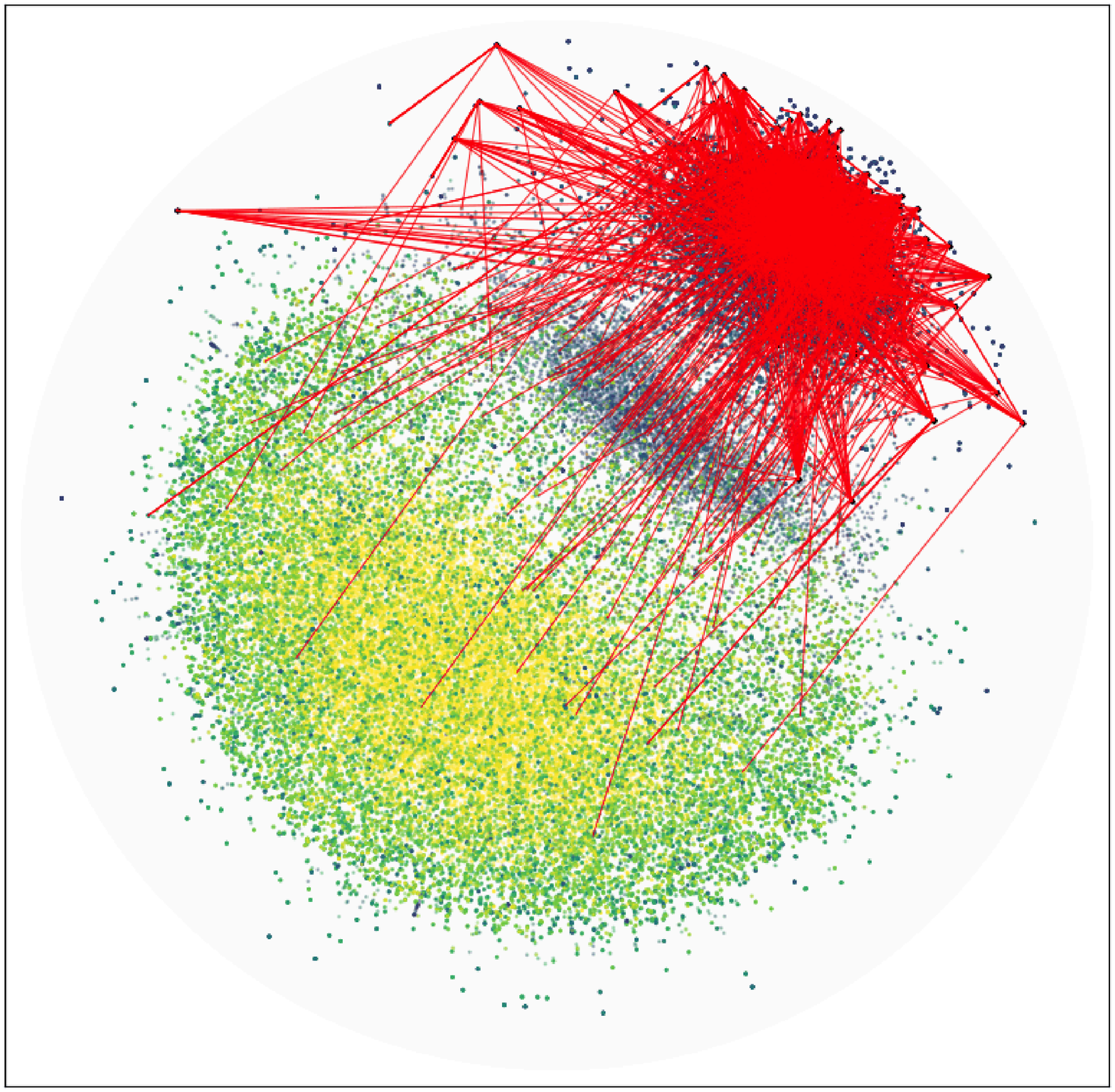} \put(0,84){d)} \put(0,94){\color{red} Links: 20} \end{overpic}
\caption{Same as Fig.~\ref{figure-learn} d). Panels highlight entities with number of their connections indicated in top left corner, starting with single entity with 32 connections in panel a) and continuing to increasing number of entities with up to 20 connections. Red lines indicate links. Upper right grouping of entities is highly inter-connected.}
\label{figure-connect}
\end{figure}

In Fig.~\ref{figure-connect} entities with highest number of connections are shown, starting with 32 and continuing to 20. Highly connected entities reside in the outer part of embeddings and are linked with upper right grouping of entities. Comparing Fig.~\ref{figure-crime} and Fig.~\ref{figure-connect} entities with positive {\tt\small fincrime\_risk\_exit} coincide with grouping of highly connected ones. 

There are 2,785 entities with positive {\tt\small fincrime\_risk\_exit} and highly connected group contains $\sim$28,000 entities, 90\% of remaining ones could be suspects for further investigation. The 3D embeddings space was chosen to give additional degree of freedom, scalability, dimensionality reduction and new pattern approach perspective (shapes) in respect to other studies using 2D embeddings, but still providing us with comprehensible insights. Entities that are ``proxies'', connecting two or more clusters of highly connected entities, in many scenarios tie together a suspicious cluster with a cluster that at the first sight looks clean. Isolated entities have less than two connections which either stand by themselves (they are flagged as initial ``unknowns'', as these do not exhibit any character or linkage for being classified as suspicious or clean), or are in the extended radius of a cluster, meaning that they will most-likely join that cluster in time and start forming relationships.

Using Poincaré embeddings it was possible to visualize and naturally classify a large interconnected datasets that would have been impossible to plot in a graph and clearly see the connections. A cluster identification technique, suitable for Poincaré distances, should be used to identify entity groupings of interest.

\section{Conclusions}
This paper demonstrated that 3D Poincaré embeddings can represent complex multi-banking customer social relations and allow identification of similarities among customers. Entities tagged by financial crime exit flag are highly connected and belong to the grouping of other highly connected entities.

The general construction of a relation high-graph is complicated and entity resolution is a major source of noise since same individuals or companies are present in registries of different banks. Working with simulated data partly addressed the problem, because individuals were uniquely identified via passport number and companies via company registration number, which simplified finding links between customers and embedding them in 3D Poincaré space.

The simulated data by SYNTHETICR aimed to follow real world financial interaction and crime scenarios to provide a global view of customer relations in multi-banking scenario. Nevertheless, consolidating the experiment in a real-bank scenario is necessary with pursue of entity resolution and deduplication before relationship embeddings to avoid noise and accuracy loss. The approach proposed in this paper could be extended to transaction data, potentially share-able as encrypted account numbers among banks, and applicable to the real world.

\section{Future work}
Going further the aim is to explore the capabilities of Poincaré embeddings in transactional relationships and to expand them by the federated learning concept using descriptive features, which are clear of personal information and can be shared between multiple banks. Analysis of known complex patterns, e.g., laundromat, and search for new patterns has to be performed. Clustering and discovery in a naturally and hierarchically embedded customer relation data is a key feature missing in current existing approaches based on matrix and graph analysis, which is needed to identify malicious networks. Carving this experiment to the financial industry domain, this paper opens the stage for near- and real- time SAR processing and data exploration, defining an architecture and topology for secure federated learning and insights sharing in a close to opaque industry.

\subsubsection*{Acknowledgments}
We thank the organizers of ``2019 Global AML and Financial Crime TechSprint'', ``Looming Threats'' team for discussions, Orestas Miskivas and Karolis Matuliauskas for their help with computing environment, and SYNTHETICR for providing the simulated data.

\bibliographystyle{unsrt}

\begin{thebibliography}{10}

\bibitem{gdpr}
{General Data Protection Regulation}.
\newblock https://eugdpr.org.

\bibitem{chicken-egg}
{Chicken and egg situation}.
\newblock https://www.collinsdictionary.com/us/dictionary/english/a-chicken-and-egg-situation.

\bibitem{aml-solutions}
{Anti-Money Laundering Solution Deep Dive}.
\newblock https://s3.amazonaws.com/cdn.ayasdi.com/wp-content/uploads/2018/04/22170635/AML\_Solutions\_Deep\_Dive\_WP\_051617v01.pdf.

\bibitem{DBLP:journals/corr/Le-KhacMOBK16}
Nhien{-}An Le{-}Khac, Sammer Markos, Michael O'Neill, Anthony Brabazon, and M.~Tahar Kechadi.
\newblock {An efficient customer search tool within an Anti-Money Laundering application implemented on an international bank's dataset}.
\newblock {\em CoRR}, abs/1609.02031, 2016.

\bibitem{DBLP:journals/corr/abs-1812-00076}
Mark Weber, Jie Chen, Toyotaro Suzumura, Aldo Pareja, Tengfei Ma, Hiroki Kanezashi, Tim Kaler, Charles~E. Leiserson, and Tao~B. Schardl.
\newblock {Scalable Graph Learning for Anti-Money Laundering: {A} First Look}.
\newblock {\em CoRR}, abs/1812.00076, 2018.

\bibitem{techsprint2019}
{2019 Global AML and Financial Crime TechSprint}.
\newblock https://www.fca.org.uk/events/techsprints/2019-global-aml-and-financial-crime-techsprint.

\bibitem{vaex}
{Vaex}.
\newblock https://vaex.io.

\bibitem{topcat}
{TOPCAT}.
\newblock http://www.star.bris.ac.uk/$\sim$mbt/topcat/.

\bibitem{DBLP:journals/corr/abs-1902-10191}
Aldo Pareja, Giacomo Domeniconi, Jie Chen, Tengfei Ma, Toyotaro Suzumura, Hiroki Kanezashi, Tim Kaler, and Charles~E. Leisersen.
\newblock {EvolveGCN: Evolving Graph Convolutional Networks for Dynamic Graphs}.
\newblock {\em CoRR}, abs/1902.10191, 2019.

\bibitem{DBLP:journals/corr/abs-1806-07464}
Stephen Bonner, Ibad Kureshi, John Brennan, Georgios Theodoropoulos, Andrew~Stephen McGough, and Boguslaw Obara.
\newblock {Exploring the Semantic Content of Unsupervised Graph Embeddings: An Empirical Study}.
\newblock {\em CoRR}, abs/1806.07464, 2018.

\bibitem{DBLP:journals/corr/abs-1709-07604}
HongYun Cai, Vincent~W. Zheng, and Kevin~Chen{-}Chuan Chang.
\newblock {A Comprehensive Survey of Graph Embedding: Problems, Techniques and Applications}.
\newblock {\em CoRR}, abs/1709.07604, 2017.

\bibitem{DBLP:journals/corr/abs-1711-08752}
Peng Cui, Xiao Wang, Jian Pei, and Wenwu Zhu.
\newblock {A Survey on Network Embedding}.
\newblock {\em CoRR}, abs/1711.08752, 2017.

\bibitem{DBLP:journals/corr/abs-1905-09791}
Ivana Balazevic, Carl Allen, and Timothy~M. Hospedales.
\newblock {Multi-relational Poincar{\'{e}} Graph Embeddings}.
\newblock {\em CoRR}, abs/1905.09791, 2019.

\bibitem{DBLP:journals/corr/NickelK17}
Maximilian Nickel and Douwe Kiela.
\newblock {Poincar{\'{e}} Embeddings for Learning Hierarchical Representations}.
\newblock {\em CoRR}, abs/1705.08039, 2017.

\end{thebibliography}
\begin{flushleft}
\begin{footnotesize}

\end{footnotesize}
\end{flushleft}

\end{document}